\documentclass{Interspeech2024}

\usepackage{graphicx}
\usepackage{multirow}
\usepackage{booktabs}
\usepackage{float}
\interspeechcameraready 
\usepackage{soul}
\usepackage{amsmath}
\usepackage{tabularx}
\DeclareTextSymbolDefault{\dh}{T1}

\newcommand{\model}{\textsc{ProEmo}}

\title{\model{}: Prompt-Driven Text-to-Speech Synthesis Based on Emotion and Intensity Control}

\name[affiliation={1}]{Shaozuo}{Zhang}
\name[affiliation={1}]{Ambuj}{Mehrish}
\name[affiliation={2}]{Yingting}{Li}
\name[affiliation={1}]{Soujanya}{Poria}

\address{
  $^1$Singapore University of Technology and Design, Singapore\\
  $^2$Beijing University of Posts and Telecommunications, China
  }
\email{shaozuo\_zhang@mymail.sutd.edu.sg, ambuj\_mehrish@sutd.edu.sg, yingting\_li@sutd.edu.sg, sporia@sutd.edu.sg }


\begin{document}
\maketitle
\begin{abstract}
 Speech synthesis has significantly advanced from statistical methods to deep neural network architectures, leading to various text-to-speech (TTS) models that closely mimic human speech patterns. However, capturing nuances such as emotion and style in speech synthesis is challenging. To address this challenge, we introduce an approach centered on prompt-based emotion control. The proposed architecture incorporates emotion and intensity control across multi-speakers. Furthermore, we leverage large language models (LLMs) to manipulate speech prosody while preserving linguistic content. Using embedding emotional cues, regulating intensity levels, and guiding prosodic variations with prompts, our approach infuses synthesized speech with human-like expressiveness and variability. Lastly, we demonstrate the effectiveness of our approach through a systematic exploration of the control mechanisms mentioned above.
\end{abstract}

\section{Introduction}
\label{sec:intro}
Speech synthesis has witnessed remarkable advances in recent years, driven primarily by the integration of deep learning techniques \cite{mehrish2023review}. While modern speech synthesis systems can produce increasingly natural-sounding speech \cite{ren2020fastspeech,kim2021conditional}, the challenge of imbuing synthesized speech with expressive qualities akin to human speech remains a focal point of research. Expressive speech synthesis \cite{govind2013expressive}, which aims to replicate the nuances of human prosody, including emotion, intonation, and speaking style, holds immense potential for various applications such as audiobook narration \cite{le2024voicebox}, virtual assistants, and conversational agents. However, achieving truly expressive speech synthesis poses significant challenges, particularly in controlling prosody to convey the desired emotional or contextual cues. 

In recent years, there has been a growing interest in expressive speech synthesis \cite{mehrish2023review,lei2023msstyletts,zhu2023multi,teh2023ensemble,triantafyllopoulos2023overview}, which aims to imbue synthesized speech with human-like prosody, emotion, and speaking style. Researchers have explored various approaches to enhance the expressiveness of synthesized speech, including prosody modeling \cite{pamisetty2023prosody, liu23n_interspeech}, emotion information embedded in speaker embeddings \cite{shaheen2023exploiting, tang23_interspeech}, and style transfer techniques \cite{wang2018style}. Additionally, the use of LLMs \cite{guo2023prompttts,sigurgeirsson2023controllable} has emerged as a promising avenue for controlling emotion in synthesized speech through prompt-based methods.

A previous research \cite{sigurgeirsson2023controllable}, has delved into the fusion of LLMs and TTS systems. They utilize prompts within LLMs to predict word-level alterations corresponding to a range of emotions, thereby facilitating the generation of expressive speech. However, their methodology lacks provisions for generating multi-speaker expressive speech. Furthermore, their approach relies solely on the output of LLMs for generating speech with varying emotions. Given that the output of LLMs can be noisy, this dependence potentially compromises the expressiveness of the generated speech. The proposed architecture \cite{sigurgeirsson2023controllable} also overlooks the incorporation of emotional intensity within its structural framework.

Generating speech with varying emotional intensity presents a significant challenge in TTS \cite{guo2023emodiff, inoue2024finegrained}. This challenge is exacerbated by the absence of explicit intensity labels in most emotional speech datasets, hindering the accurate conveyance of nuanced emotions. Emotion intensity, being highly subjective and complex, poses a greater challenge compared to discrete emotion categorization. Existing approaches in the literature mainly fall into two categories for controlling emotion intensity in TTS. One approach involves incorporating auxiliary features such as the state of voiced, unvoiced, and silence (VUS), attention weights, or saliency maps. The other approach manipulates internal representations of emotion through techniques like interpolation or scaling. Despite these efforts, effectively controlling emotion intensity in TTS remains an underexplored area. 
\begin{figure*}
    \centering
    \includegraphics[width=1.7\columnwidth]{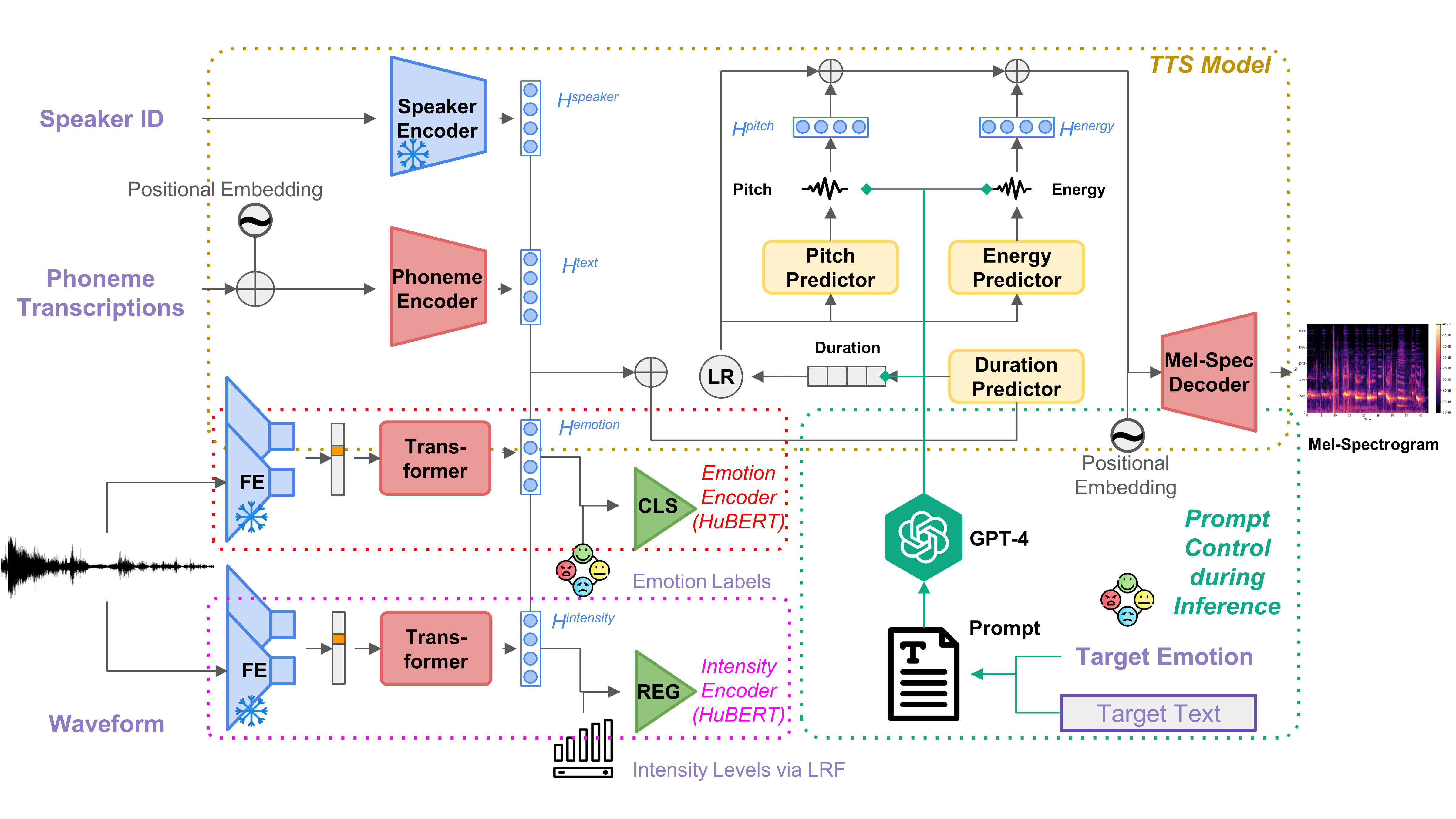}
    \vspace{-5mm}
    \caption{Overview of proposed expressive speech generation framework composed of $4$ modules: TTS backbone based on FS2(yellow-dotted), HuBERT for Emotion Encoder(red-dotted), HuBERT for Intensity Encoder(purple-dotted), GPT-4 Prompting for prosody control(green-dotted). \textbf{FE}: Feature Extractor, \textbf{CLS}: Classification Head, \textbf{REG}: Regression Head. \textbf{LRF}: Learned Rank Function}
    \label{fig:proemo_arc}
    \vspace{-5mm}
\end{figure*}

In this paper, we propose integrating emotion and intensity encoders into the FastSpeech 2 (FS2) architecture to enhance the expressiveness and versatility of synthesized speech. Additionally, we leverage the capabilities of LLMs (GPT-4) to predict prosody and modulate emotions in speech during inference. Our approach represents a departure from traditional methods of emotion manipulation, offering a promising avenue for enhancing the expressiveness and naturalness of synthesized speech. Our main contributions are summarized as follows:

\begin{itemize}
    \item We propose a novel TTS framework that extends FS2 for prompt-based prosody prediction, enabling emotion manipulation during inference.
    \item We successfully control pitch, energy and duration in both utterance and word level via the novel LLM prompt. 
    \item Our framework can generate multi-speaker expressive speech with varying emotional intensity.
\end{itemize}



 
\section{Related Works}
\label{sec:rw}
Recent TTS advancements explore specifying target prosody, emotion, or speaking style through natural language prompts. PromptTTS \cite{guo2023prompttts} introduces a style encoder trained on 'style prompts', natural language descriptions of desired speaking styles. This encoder predicts values for categorical parameters (e.g., gender and emotion) to guide speech generation, akin to models like InstructGPT \cite{ouyang2022training}. While simplifying style specification, this approach heavily relies on scarce training data labeled with ground-truth style prompts. InstructTTS \cite{yang2023instructtts} also employs prompt-based control by generating a latent speaking-style representation from speech, text, and a ground-truth style prompt. This representation conditions both the text encoder and a diffusion-based decoder allowing for defining new speaking styles via natural language but requiring a corpus annotated with style prompts for effective training.

\section{Methodology}
\label{sec:med}
The proposed pipeline is structured into distinct stages. Initially, we pre-train a multispeaker English TTS backbone model utilizing a large publicly accessible dataset. Then, we add emotion and intensity encoders and fine-tune the model with an emotional dataset. During inference, we use a LLM to predict prosody aligned with the target emotion and make subtle adjustments in pitch, duration, and energy for desired effects.
\vspace{-0.5em}
\subsection{Text-to-Speech Backbone}
Our TTS system employs FS2 \cite{ren2020fastspeech} as its backbone, comprising a phoneme encoder, a speaker encoder, a variance adaptor, and a mel-spectrogram decoder. The Mel-Spectrogram is converted into a waveform using a pre-trained HiFiGAN \cite{kong2020hifigan} vocoder. FS 2 is chosen for its interpretable prosody modification capability: during inference, predicted pitch, energy, and duration can be adjusted to meet specific prosodic requirements.
\vspace{-0.5em}
\subsection{Emotion Encoder}
\label{subsec:ee}
\vspace{-0.5em}
FS2 enhances speed and voice quality, yet its architecture lacks inherent emotional expression, hindering expressive speech generation. To address this, we integrate an emotion encoder to enhance its expressiveness. The emotion encoder learns emotion representations directly from the waveform, utilizing Hubert \cite{hsu2021hubert} for feature extraction and a classification head for emotion recognition. During training, only the transformer layer of Hubert and the classification head are fine-tuned, while the convolutional layers remain frozen. The resulting emotion embeddings seamlessly integrate with the backbone TTS model. This integrated architecture is denoted as FS2$_{w/ Emo}$.

\subsection{Intensity Encoder}
\label{subsec:ie}
Expressive speech relies heavily on intensity to convey emotions sincerely. To enhance synthesized speech with nuanced emotions, we introduce an intensity encoder. Similar to the emotion encoder, it utilizes Hubert \cite{hsu2021hubert} and a regression head to predict emotion intensity levels. Modeling emotion intensity becomes crucial as most emotional speech datasets lack intensity annotation. Inspired by \cite{zhou2022emotion}, we view emotion intensity as a fundamental attribute of the emotional speech. However, annotating intensity labels poses challenges, even for humans \cite{ferrari2007learning}. \cite{zhou2022emotion} addresses this by using relative attribute labeling, similar to its success in computer vision \cite{ferrari2007learning}. We intend to improve upon the algorithm proposed in \cite{zhou2022emotion} for predicting emotion intensity levels using a multispeaker emotive speech dataset.

We quantify intensity by comparing emotional and neutral speeches using acoustic features. We utilize an open-source implementation to train a relative ranking function for emotion intensity, denoted as $r(x_A) = \textbf{W}x_A$. Here, $x_A$ represents $384$-dimensional acoustic features extracted from openSMILE \cite{eyben2010opensmile}, \textbf{W} denotes trainable weighting parameters, and $r(x_A)$ normalizes intensity values from 0 to 1. Departing from the prior work, we propose learning $r(x_A)$ for each speaker $A$ in the training set $T$, allowing nuanced modeling of intensity variations tailored to individual speakers. We also develop distinct ranking functions for each emotion category, leveraging both neural and emotional utterances. During training, the weighting matrix is computed akin to an SVM problem, detailed in \cite{parikh2011relative}. This learned function enables scoring new data based on its acoustic features. As intensity is continuous, the intensity encoder's training is supervised through a regression task. We denote the TTS architecture with the intensity encoder as FS2$_{w/ Emo\&Int}$.

\vspace{-0.5em}
\subsection{Prompt Control}
\label{subsec:pc}
\begin{figure}[ht]
  \centering
  \includegraphics[width=0.9\columnwidth]{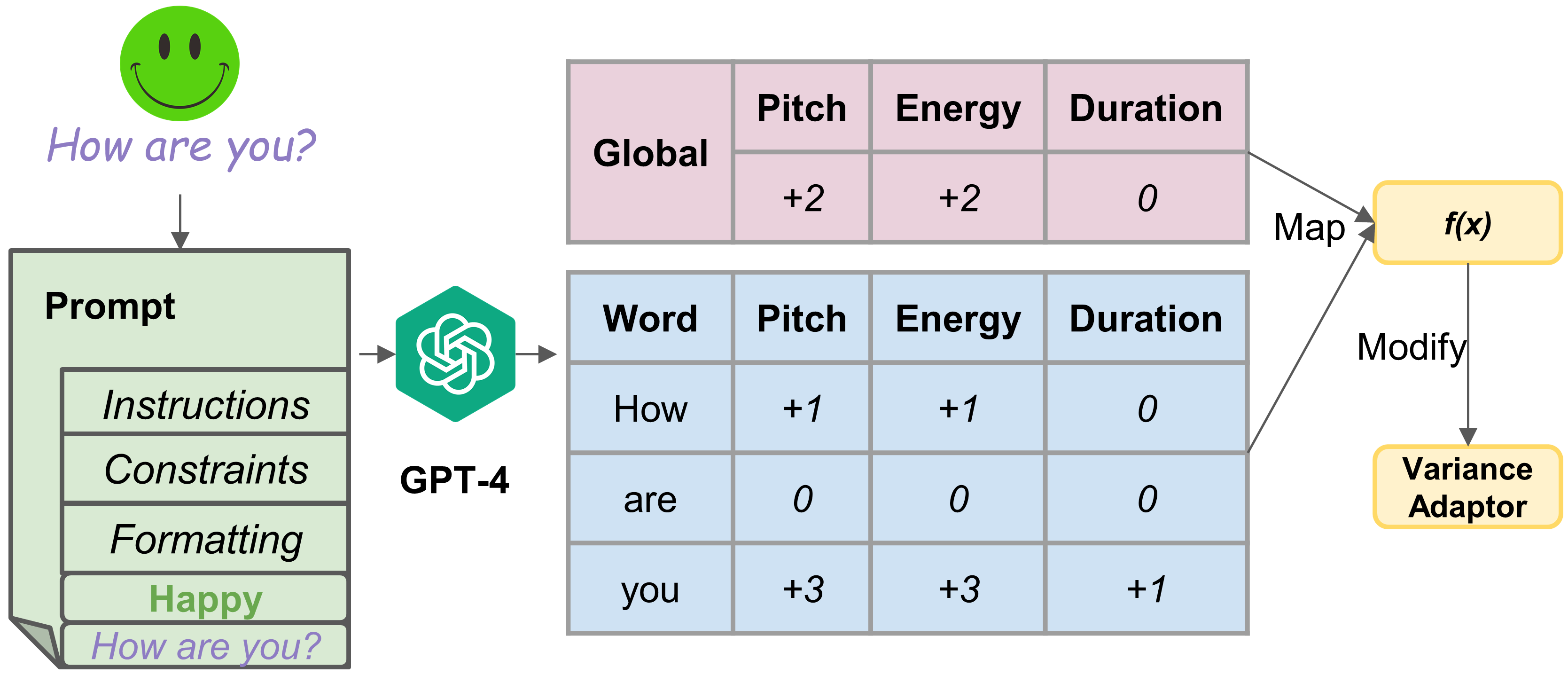} 
  \caption{Introduction to Prompt Control: The scaling factors suggested by the LLM (shown in red and blue tables) directly affect the Variance Adaptor}
  \label{fig:prompt_control}
\end{figure}

During inference, the FS2 predicts duration ($d$), energy ($e$), and pitch ($p$) per phoneme based on the input text. Inspired by \cite{sigurgeirsson2023controllable}, we aim to enhance speech generation to match desired emotional tones by adjusting $d$, $p$, and $e$ values per phoneme. Following a similar method, we adopt a two-level modification approach to regulate prosodic effects at both the utterance ("global") and word ("local") levels sequentially. The global modification establishes the overall emotional tone, while subsequent local modifications refine voice variances between individual words for a more nuanced, emotionally expressive output. We term this as "prompt control," illustrated in Figure \ref{fig:prompt_control}.

Similar to \cite{sigurgeirsson2023controllable}, we mathematically formulate the modification process with Equations \ref{eq:dur}, \ref{eq:energy}, and \ref{eq:pitch}. Let $i$ denote the $i$-th word in the utterance, with predicted values by the variance adaptor as $d_i$, $e_i$, and $p_i$ for duration, energy, and pitch respectively. Their scaled counterparts are denoted as $d^\prime_i$, $e^\prime_i$, and $p^\prime_i$. Additionally, $G_d$, $G_e$, and $G_p$ indicate the global scaling factors for duration, energy, and pitch respectively, while $\sigma_j$, $\epsilon_j$, and $\pi_i$ represent the corresponding local scaling factors.

\vspace{-5mm}
\begin{align}
    &d^\prime_i = d_i \cdot G_d \cdot \sigma_i, &G_d, \sigma_i &\in [0.74, 1.34] \label{eq:dur} \\
    &e^\prime_i = e_i \cdot G_e \cdot \epsilon_i, &G_e, \epsilon_i &\in [0.5, 2] \label{eq:energy} \\
    &p^\prime_i = p_i + G_p + \pi_i, &G_p + \pi_i &\in [p_{\text{min}}, p_{\text{max}}] \label{eq:pitch}
\end{align}


where $p_{min}$ and $p_{max}$ denote the minimum and maximum pitch values predicted by the FS2 variance adaptor. For both global and local scaling factors related to duration and energy, we confine the range between $[0.5, 2]$, equivalent to halving and doubling the duration and energy, respectively, to preserve naturalness in speech \cite{sigurgeirsson2023controllable}. However, our experimental observations reveal that scaling duration is more sensitive than scaling energy. Consequently, we empirically narrowed the interval for $G_d$ and $\sigma_i$ to $[0.74, 1.34]$. We direct GPT-4 to predict both global and local scaling factors within $[-5, 5]$ for pitch and energy, and $[-2, 2]$ for duration. Subsequently, we map the predicted values using the quadratic map function $f(x) = ax^2 + bx + c$\footnote{ where $a$, $b$, and $c$ are estimated using defined boundary condition of $[0.5, 2]$.} to intervals specified in Eq. \ref{eq:dur}, \ref{eq:energy}, and \ref{eq:pitch}.
 

The prompt design is a crucial aspect of our methodology. Initially, we tried a prompt from \cite{sigurgeirsson2023controllable} but faced instability in GPT-4 output. This instability caused consistent changes in pitch, energy, and duration, leading to poor expressiveness. To tackle this, we created a comprehensive prompt template ( Figure \ref{fig:prompt_control}) covering task description and output requirements. We then used the GPT-4 model via the OpenAI API \cite{gpt4} to obtain appropriate scaling factors. Our prompt encourages reasoning at each decision step to reduce potential noise.

\begin{figure*}[ht]
\centering
    \includegraphics[width=0.8\linewidth]{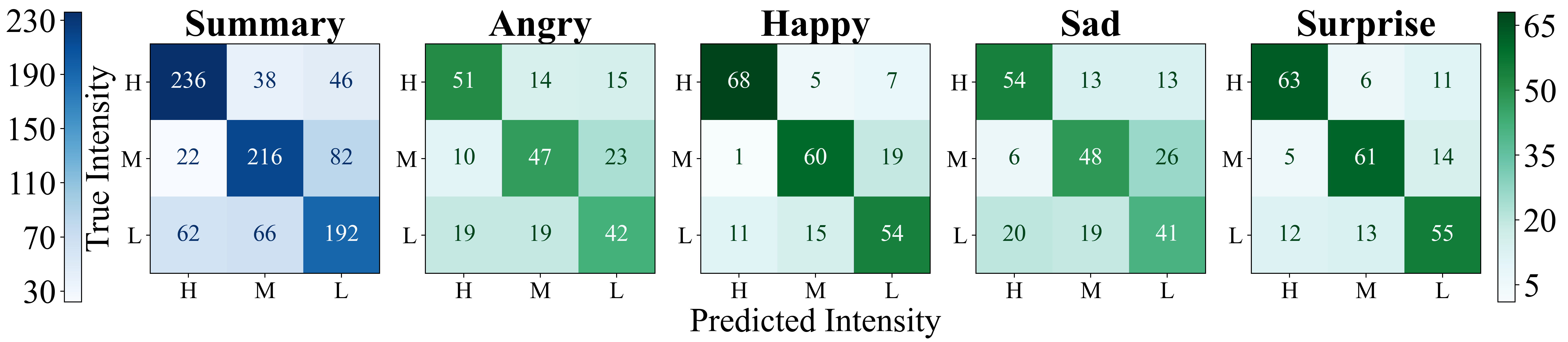}
    \caption{Perceptual Intensity Ranking. \textbf{H,M,L}: High, Medium and Low intensities.}
    \label{fig:int_cms}
    \vspace{-5mm}
\end{figure*}

\vspace{-0.5em}
\section{Experimental Setup}
\subsection{Baselines and Datasets}
We utilize the LibriTTS \cite{zen2019libritts} for pretraining, which includes $33,236$ training samples (equating to $53.78$ hours) collected from $247$ speakers. To capture various emotions, we turn to the Emotional Speech Database (ESD) for fine-tuning. This dataset consists of $350$ utterances, totaling approximately 13.4 hours of recordings, by $10$ English speakers, and encompasses five emotion categories: Angry, Happy, Neutral, Sad, and Surprise.

For expressive speech synthesis, we utilize two established architectures: Daft-Exprt \cite{zaidi2021daft} and FS2. Following the approach in \cite{zaidi2021daft-git}, we pre-train the Daft-Exprt model on LJSpeech before fine-tuning it on the ESD. Despite attempts to pre-train on LibriTTS, it failes to converge, resulting in poor-quality speech. Hence, we stick to the method proposed in \cite{zaidi2021daft} for training. We experiment with different FS2 configurations: 1) FS2, 2) FS2$_{w/ Emo}$ with an emotion encoder, and 3) FS2$_{w/ Emo\&Int}$ with both emotion and intensity encoders. During inference, we maintain consistency by applying the prompt control discussed in Section \ref{subsec:pc} across all models for a uniform evaluation.

\vspace{-0.5em}
\subsection{Training and Evaluation}
\label{subsec:trainandeval}
\vspace{-0.5em}
The training process involves two main stages: initial pre-training on a substantial English corpus, followed by fine-tuning on an emotional corpus with auxiliary supervision tasks. In the pre-training phase, multispeaker FS2 training is conducted on the LibriTTS-100 for 900K steps, with a batch size of $16$. Speaker embeddings are computed using a speaker verification model trained with the GE2E loss \cite{wan2018generalized}.



In the subsequent stage of training, we refine the pre-trained multispeaker FS2 model on the ESD corpus by incorporating emotion and intensity encoders as discussed in section \ref{sec:med}. These encoders, based on the HuBERT \cite{hsu2021hubert-hf}, are trained with the backbone FS2 model in an end-to-end fashion. During training, the feature extractor module of HuBERT remains fixed, while only the transformer parameters, along with the classification/regression head parameters, are updated for $50$K steps on the ESD dataset. We train different versions of FS2, each detailed in Table \ref{tbl:performance_table}. During inference, prosody control is achieved using GPT4, as elaborated in Section \ref{subsec:pc}\footnote{Source codes, the prompt template with examples and audio samples are available at \url{https://anonymous.4open.science/r/FastSpeech2EmoInt-072F}.}. For evaluation, we use the ESD test set with four distinct emotions. To compute objective metrics, we generate expressive speech under three distinct prompt control settings: (1) without any prompt control; (2) with both global and local level prompt controls; and (3) exclusively with local level control.



To assess our methodology's effectiveness, we employ a comprehensive framework with various metrics. These include emotion classification accuracy (ECA) to gauge prosody control's impact on speech expressiveness. We also use Mel Cepstral Distortion (MCD) \cite{kominek2008synthesizer} to measure speech distortion. Word Error Rate (WER) and Character Error Rate (CER) \cite{radford2023robust} to evaluate alignment with ground truth. Additionally, Mean Opinion Score (MOS) tests \cite{streijl2016mos} involve human listeners rating synthesized speech quality numerically.

Furthermore, we follow the methodology outlined in \cite{wang2023fine} the Perceptual Intensity Ranking (PIR) test, for subjectively comparing synthesized speech samples with different intensity levels. In our experiments, we generate speech samples at three intensity levels: Low, Medium, and High. Since ground truth intensity annotations are lacking, we derive them using the learned ranked function, detailed in Section \ref{subsec:ie}, for the evaluation dataset. These annotations serve as a reference during the PIR test. Participants rank generated samples based on perceived intensity, categorizing them into predefined levels, which are then compared with intensity annotations derived from the learned rank function.

\section{Results}
\subsection{Objective Evaluation}
\begin{figure}[ht]
  \centering
  \includegraphics[width=\columnwidth]{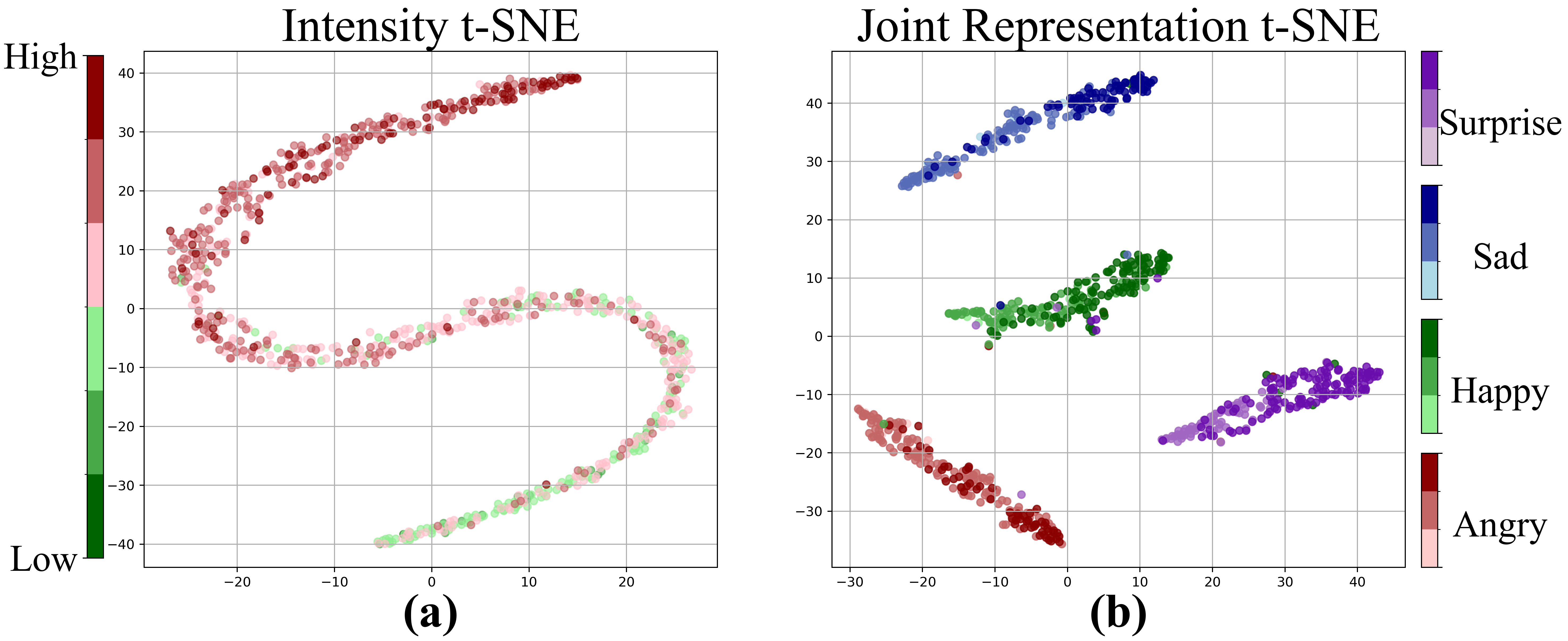}
  \vspace{-5mm}
  \caption{t-SNE of embeddings for the ESD validation set. (a) Intensity embeddings computed using the relative ranking function, $r(.)$. (b) Joint emotion and intensity embeddings.}
  \label{fig:dual_tsne}
  \vspace{-6mm}
\end{figure}

To assess synthesized speech expressiveness, we analyze emotion recognition across different models. We train an emotion encoder independently using the ESD training set. This encoder achieves $95\%$ accuracy on the ESD evaluation set, showcasing its ability to capture subtle emotional nuances. Further insights into emotion representation quality and intensity are depicted in Fig. \ref{fig:dual_tsne}, illustrating the encoders' discriminative ability in recognizing emotions and gauging intensity with acceptable deviation. We then use this trained emotion encoder to classify emotions in speech generated by different models, as shown in Table \ref{tbl:performance_table}.
\vspace{-3mm}


\begin{table}[h!t]
\centering
\small
\setlength{\tabcolsep}{2.85pt}
\begin{tabular}{@{}lllllll@{}}
\toprule
    \textbf{Model} & \textbf{PC} & \textbf{ECA} $\uparrow$ & \textbf{MCD} $\downarrow$ & \textbf{WER} $\downarrow$ & \textbf{CER} $\downarrow$ & MOS $\uparrow$ \\ \midrule
    \textit{Daft-Exprt}& None & 0.663 & \textbf{6.278} & 0.353 & 0.123 & 3.450 \\
    \textit{Daft-Exprt}& G\&L & 0.683 & 6.659 & 0.657 & 0.386 & --\\
    \textit{Daft-Exprt}& L & 0.481 & 5.889 & 0.969 & 0.833 & -- \\
    \midrule
    FS2 & None & 0.247 & 6.710 & 0.125 & 0.040 & -- \\
    FS2 & G\&L & 0.297 & 7.160 & 0.135 & 0.049 & -- \\ 
    FS2 & L & 0.296 & 6.865 & 0.129 & 0.042 & -- \\ 
    \midrule
    FS2$_{w/E}$ & None & 0.762 & 6.619 & \textbf{0.127} & 0.039 & --\\ 
    FS2$_{w/E}$ & G\&L & 0.782 & 6.968 & 0.131 & 0.044 & --\\ 
    FS2$_{w/E}$ & L & \textbf{0.820} & 6.767 & 0.129 & 0.039 & --\\ \midrule
    FS2$_{w/E\&I}$ & None & 0.748 & 6.594 & 0.125 & \textbf{0.038} & 3.194\\ 
    FS2$_{w/E\&I}$ & G\&L & 0.771 & 7.058 & 0.130 & 0.043 & \textbf{3.728}\\ 
    FS2$_{w/E\&I}$ & L & \textbf{0.797} & 6.741 & 0.128 & 0.040 & 3.408\\ 
\bottomrule
\end{tabular}
\vspace{1mm}
\caption{Objective \& Subjective Evaluation Performance. \textbf{PC}: Prompt Control; \textbf{ECA}: Emotion Classification Accuracy; \textbf{FS2}: FastSpeech 2; \textbf{G\&L}: Both global \& local level prompt control; \textbf{L}: Local level prompt control}.
\label{tbl:performance_table}
\vspace{-5mm}
\end{table}

Table \ref{tbl:performance_table} reveals a consistent trend: incorporating local-level modification during inference consistently leads to higher emotion classification accuracy compared to both global and local-level control, as well as instances without any prompt control. For instance, in the case of FS2$_{w/ Emo \& Int}$, classification accuracy improves to $79.72\%$ with local-level control compared to $74.80\%$ without any prompt control. This trend persists across both FS2$_{w/ Emo \& Int}$ and FS2$_{w/ Emo}$ models, highlighting the significant enhancement in speech expressiveness through prompt-level control. Additionally, analysis of MCD, WER, and CER scores from Table \ref{tbl:performance_table} suggests that the inclusion of emotion and intensity embeddings, alongside prompt control, does not significantly affect speech generation. Moreover, the lower performance of Daft-Exprt compared to FastSpeech2 can be attributed to the former's pretraining on a single-speaker dataset, limiting its ability to adapt to various prosodic changes inherent in expressive speech synthesis.
\vspace{-2mm}
\subsection{Human Subjective Study}
\vspace{-1mm}
 We invite $20$ participants (7 females, 13 males) for a human subjective study. These participants from diverse geographical regions are expertise in speech and NLP. 


\noindent\textbf{Mean Opinion Score(MOS) Test}
We instruct GPT-4 to generate five sentences randomly for each emotion category. Then, we synthesize speech based on three different prompt control settings detailed in Section \ref{subsec:pc}. Participants evaluate each audio clip based on expressiveness and naturalness, using a rating scale from $0$ to $5$. The MOS results in Table \ref{tbl:performance_table} reveal that the FS2$_{w/ Emo \& Int}$ configuration with global \& local level prompt control outperforms others across all generated emotions.

\noindent\textbf{PIR Test:} We randomly select four utterances per emotion category from the ESD evaluation set and reconstruct them at three intensity levels (Low, Medium, and High) using intensity annotations from the learned rank function (Section \ref{subsec:trainandeval}). This yields a total of $16 \times 3 = 48$ samples for evaluation. Participants rank these audio samples based on perceived intensity. Results in Figure \ref{fig:int_cms} show confusion matrices for FS2$_{w/ E \& I}$ with word-level prompt control. Overall, participants achieve approximately $72\%$ accuracy in perceiving correct intensity levels, indicating the effectiveness of our method in generating speech with varied intensity.
\vspace{-2mm}

\section{Conclusion}
\vspace{-1mm}
In conclusion, our proposed approach, focused on prompt-based emotion control in speech synthesis, represents a significant advancement in infusing synthesized speech with human-like expressiveness and variability. By integrating emotion and intensity control across multi-speakers and utilizing large LLMs to manipulate speech prosody while preserving linguistic content, we've presented a novel framework that enhances the naturalness and versatility of synthesized speech. Our systematic exploration of control mechanisms has yielded promising results in generating expressive speech with varying emotional intensity. Moving forward, future research and development opportunities include considerations for Multilingual and Cross-cultural contexts, enhancing Robustness and Generalization, and exploring Real-time Applications in virtual assistants or live communications platforms.

\bibliographystyle{IEEEtran}
\bibliography{mybib}

\end{document}